**Notes on the UK Non-Native Organism Risk Assessment Scheme**[1]

Gareth Hughes[2]
School of Biological Sciences, University of Edinburgh, Edinburgh EH9 3JG, U.K.

In 2004, the UK Government's Department for Environment, Food and Rural Affairs (DEFRA) commissioned research with the aim of developing a scheme for assessing the risks posed to species, habitats and ecosystems in the UK by non-native organisms. The outcome was the UK Non-Native Organism Risk Assessment Scheme. A comprehensive overview of the risk assessment methodology and its application is provided by Baker et al. (2008). Unfortunately, the mathematical basis of the procedure for summarising risks deployed in the Risk Assessment Scheme, as outlined in Baker et al. (2008, section 3.5) and described in more detail in Module 5 of the Risk Assessment Scheme's User Manual[3] contains several analytical errors. These errors are outlined in the notes that follow.

**Background**

The Risk Assessment Scheme is a decision-making apparatus, based on the application of Bayes' theorem. The general objective is to provide a basis for discrimination between species likely to cause environmental problems if imported (referred to as *invasive*), and those unlikely so to do (referred to as *non-invasive*). It is not a practical proposition to discriminate by allowing entry and recording the consequences, then taking action as necessary. The specific objective of the analysis is therefore the development of a predictor (i.e., a diagnostic test) for invasiveness. An *indicator variable* (or, for brevity, an indicator) is a proxy variable related to the actual variable of interest (in this case, invasiveness), used because it is difficult to obtain directly data relating to the actual variable of interest. The first step is the characterization of one or more useful indicators. In the Risk Assessment Scheme there are 51 *risk components* in four major categories (entry, establishment, spread and impact) that serve as indicators (although not all of them are necessarily used in any individual assessment).

To facilitate an explanation of what is wrong with the analysis on which the Risk Assessment Scheme is based, it will be helpful first to see briefly an outline of what a correct analysis should look like. The generic methodology for the development of predictors has been applied to great effect in clinical epidemiology. Somoza et al. (1989) provide a useful summary. In an epidemiological context, two mutually exclusive groups of subjects are identified, one of individuals designated *cases*, definitively having disease status $D_1$, the other of individuals designated *controls*, definitively having disease status $D_2$. This definitive identification is often referred to as the *gold standard*. The gold standard classification into case and control groups is made independent of the putative indicator.

---

[1] These notes © Gareth Hughes 2008
[2] email: g.hughes@ed.ac.uk
[3] http://collections.europarchive.org/tna/20080906003302/http://www.defra.gov.uk/wildlife-countryside/resprog/findings/non-native-risks/index.htm

A 5-point ordinal scale of rating categories is often used to assess the status of subjects using the indicator. Each point on the rating category scale corresponds to a score $s_i$ ($i$ = 1,2,…,5). The indicator is usually calibrated in such a way that cases tend to have larger scores than controls. The indicator scores are recorded for all subjects in both groups, and the resulting frequency distributions of scores tabulated separately for cases and controls. Generally, the two frequency distributions of indicator scores overlap. Table 1 shows an example taken from Swets (1988). The data are taken from a study in which six radiologists attempted independently to distinguish between malignant and benign lesions as viewed in a set of 118 mammograms. The gold standard classification (independent of the radiologists' inspections of the mammograms, of course) showed 58 malignant and 60 benign lesions, so there are, in this example, 6×(58+60)=708 subjects (mammogram inspections) comprising 348 cases and 360 controls.

Table 1. Mammography data from Swets (1988)

| Rating category | Frequency data | | Probability | | Likelihood ratio |
|---|---|---|---|---|---|
|  | Cases | Controls | Cases | Controls |  |
| Very likely benign | 15 | 92 | 0.043 | 0.256 | 0.17 |
| Probably benign | 53 | 151 | 0.152 | 0.419 | 0.36 |
| Possibly malignant | 63 | 48 | 0.181 | 0.133 | 1.36 |
| Probably malignant | 85 | 50 | 0.244 | 0.139 | 1.76 |
| Very likely malignant | 132 | 19 | 0.379 | 0.053 | 7.19 |
| Total | 348 | 360 | 1.000 | 1.000 |  |

The probabilities and likelihood ratios in Table 1 are shown in Table 2 in notational terms. The probabilities are conditional probabilities: we read $P(s_i=5|D_1)$ as the probability that a subject will be rated as $s_i=5$ given that the subject is a case, and $P(s_i=5|D_2)$ as the probability that a subject will be rated as $s_i=5$ given that the subject is a control. The likelihood ratios are the ratios of conditional probabilities for corresponding scores (Table 2). The likelihood ratio $L(s_i=5) = 7.19$ (Table 1) means that scores of $s_i=5$ are just over seven times more likely to come from subjects that are cases than from subjects that are controls.

Table 2. Notation for mammography data from Swets (1988)

| Rating category, $i$ | Score, $s_i$ | Probability, P | | Likelihood ratio, $L(s_i)$ |
|---|---|---|---|---|
|  |  | Cases | Controls |  |
| 1 | 1 | $P(s_i=1|D_1)$ | $P(s_i=1|D_2)$ | $P(s_i=1|D_1)/P(s_i=1|D_2)$ |
| 2 | 2 | $P(s_i=2|D_1)$ | $P(s_i=2|D_2)$ | $P(s_i=2|D_1)/P(s_i=2|D_2)$ |
| 3 | 3 | $P(s_i=3|D_1)$ | $P(s_i=3|D_2)$ | $P(s_i=3|D_1)/P(s_i=3|D_2)$ |
| 4 | 4 | $P(s_i=4|D_1)$ | $P(s_i=4|D_2)$ | $P(s_i=4|D_1)/P(s_i=4|D_2)$ |
| 5 | 5 | $P(s_i=5|D_1)$ | $P(s_i=5|D_2)$ | $P(s_i=5|D_1)/P(s_i=5|D_2)$ |
| Total |  | 1.000 | 1.000 |  |

All this is useful because it allows us to make evidence-based revisions of the probability of case status. Once we have carried out the diagnostic test on a subject of unknown status, we know the rating category score for the indicator variable (i.e., risk component) and so the corresponding likelihood ratio. The likelihood ratio is combined with information on the *prevalence* of cases in the population of subjects of interest, using Bayes' theorem, in order to calculate the conditional probability of case status, given the evidence obtained from the application of the diagnostic test. In more detail, we proceed as follows.

1. Think of the prevalence as the prior (i.e., pre-test) probability of case status in the population of interest, denoted $P(D_1)$. In general, the proportion of cases in the experimental data will not be an appropriate estimate of $P(D_1)$. Such an estimate would be made independent of the experimental data.
2. Calculate the prior odds $O(D_1) = P(D_1)/(1 - P(D_1))$.
3. Carry out the diagnostic test. This tells us the score $s_i$ for the subject under consideration. Thus we can obtain the corresponding likelihood ratio $L(s_i)$.
4. Using Bayes' theorem, calculate the posterior odds $O(D_1|s_i) = O(D_1) \times L(s_i)$.
5. Calculate the posterior probability $P(D_1|s_i) = O(D_1|s_i)/(1 + O(D_1|s_i))$.

More generally, we may have data relating more than one risk component. Suppose, then, that a risk assessment will be based on a number (denoted $n$, $n \geq 1$) of *independent* risk components. Each risk component supplies a diagnostic test. We will assume (as in the Risk Assessment Scheme) that each of the risk components is assessed on an ordinal categorical 5-point scale. We then proceed as follows.

1. Think of the prevalence as the prior (i.e., pre-test) probability of case status in the population of interest, denoted $P(D_1)$. As above, an estimate of $P(D_1)$ would be made independent of the experimental data.
2. Calculate the prior odds $O(D_1) = P(D_1)/(1 - P(D_1))$.
3. Carry out the diagnostic test(s). This tells us the score $s_i$ for the subject under consideration for each of the $n$ risk components. Thus we can obtain a set of $n$ corresponding likelihood ratios $L(s_i)_1 \ldots L(s_i)_n$.
4. A combined likelihood ratio is given by the product of the individual likelihood ratios for the diagnostic tests:

$$\prod_n L\{s_i\} = L(s_i)_1 \times L(s_i)_2 \times \cdots \times L(s_i)_n$$

   (Go, 1998), in which $\{s_i\}$ denotes the set of all the scores for the $n$ risk components.
5. Using Bayes' theorem, calculate the posterior odds:

$$O(D_1|\{s_i\}) = O(D_1) \times \prod_n L\{s_i\}.$$

6. Calculate the posterior probability $P(D_1|\{s_i\}) = O(D_1|\{s_i\})/(1 + O(D_1|\{s_i\}))$.

$P(D_1|\{s_i\})$ is the probability of case status given the outcome of the diagnostic test(s). Thus we have revised our prior probability $P(D_1)$, taking into account evidence related to one or more risk components. The essential idea is that the posterior probability $P(D_1|\{s_i\})$ provides an improvement over the prior probability $P(D_1)$ as a basis for evidence-based decision-making.

**The Risk Assessment Scheme**

Here, for brevity, we consider only the overall risk assessment based on up to 51 indicators (referred to as risk components) in four categories (the same methodological problems arise if the four categories of risk are analysed separately). The notation used is as follows: "$v$" – invasive (cases); "$\neg v$" – non-invasive (controls).

1. Score-to-probability conversion

The Risk Assessment User Manual says:

"*Let the probability associated with a score of '2' be 0.5. The extent to which the probability is greater or less than 0.5 when scores are greater or less than '2' is defined by the score to probability conversion parameter. This defines the increment in probability terms for each score point increment.*"

The resulting score-to-probability conversions are shown in Table 3. In this case, the scores $s_i$=0,1,2,3,4 are given to rating categories $i$=1,2,3,4,5.

Table 3. Score-to-probability conversion for the Risk Assessment Scheme

| Rating category, $i$ | Score*, $s_i$ | Probability | Probability calculation** |
|---|---|---|---|
| 1 | 0 | 0.466 | $= 0.5 - (2 \times 0.017)$ |
| 2 | 1 | 0.483 | $= 0.5 - (1 \times 0.017)$ |
| 3 | 2 | 0.500 | $= 0.5$ |
| 4 | 3 | 0.517 | $= 0.5 + (1 \times 0.017)$ |
| 5 | 4 | 0.534 | $= 0.5 + (2 \times 0.017)$ |

*0 – the lowest risk to 4 – the highest risk
** 0.017 is an empirical score-to-probability conversion factor

Recall that in the mammography experiment, two data values were recorded from each individual subject: (1) the gold standard classification (case or control), and (2) a radiologist's rating category score. Following the experiment, we could say, for example, that an estimate of $P(s_i=3|D_1)$ is 0.181 because 63 out of 348 cases were rated as $s_i$=3. Similarly, an estimate of $P(s_i=3|D_2)$ is 0.133 because 48 out of 360 controls were rated as $s_i$=3. These (conditional) probabilities are based on relative frequencies of occurrence of an outcome in a well-defined experiment. They provide a basis for the calculation of the likelihood ratio $L(s_i=3)$ and the application of Bayes' theorem (once we have an estimate of the prior probability) to calculate the posterior probability $P(D_1|s_i=3)$.

The probabilities given in Table 3 are clearly different in nature to those obtained in the mammography experiment (Tables 1 & 2). How are they to be interpreted?

The Risk Assessment User Manual says:

"*The set of starting probabilities are defined as the conditional probabilities that an organism is invasive given that it has a particular score for a particular question*[4]."

That is to say, the calculated probabilities in Table 3 are taken to be estimates of the posterior probabilities $P(v|s_i)$ (equivalent to $P(D_1|s_i)$ in the mammography experiment). So, for example, according to this interpretation, if an organism (being a sample from a particular species under consideration) is rated as $s_i=2$ for a particular risk component, the probability that the species represented by the organism under assessment is invasive, based on the score for the particular risk component in question, is $P(v|s_i)=0.5$. Such probabilities represent the degree of belief (in this case, by the authors of the Risk Assessment Scheme) in the statement that the particular species under consideration is invasive. Of course, the authors of the Risk Assessment Scheme are perfectly entitled to express their beliefs in probabilistic terms in this way.

2. Calculation of conditional probabilities

Unfortunately, things now start to go seriously wrong.

The Risk Assessment User Manual says:

"*We define two probabilities*:

$P(s_i|v)$ = *probability that risk component i is given a certain score, given that the species concerned poses an invasion risk.*

$P(s_i|\neg v)$ = *probability that risk component i is given a certain score, given that the species concerned does NOT pose an invasion risk.*

*It is assumed that* $P(s_i|\neg v) + P(s_i|v) = 1$."

Here, the probabilities in Table 3 are defined as the conditional probabilities $P(s_i|v)$ (equivalent to $P(s_i|D_1)$ in the mammography experiment). This is in apparent contradiction to the previous statement quoted above (i.e., the conditionality has been reversed), but we will return to this point later. The first problem here is that we are not actually given that any species poses an invasion risk because, unlike the mammography experiment, there is no gold standard classification of cases (and controls). The introduction of conditionality implies that we have some *additional information* about an event that might influence our view of its probability (or odds). The analysis on which the Risk Assessment Scheme is based provides information on scores for risk components but not on the definitive status of species, invasive or non-invasive. A second problem arises with the derivation of the $P(s_i|\neg v)$ (equivalent to $P(s_i|D_2)$ in the mammography experiment). It is assumed that $P(s_i|\neg v) + P(s_i|v) = 1$, so $P(s_i|\neg v) = 1 - P(s_i|v)$, but "$v$" and "$\neg v$" represent two different sample spaces so the assumption cannot, in general, be valid. See, for example, Table 1, where $P(s_i|D_2) + P(s_i|D_1) \neq 1$ for any *i*. Thirdly, for the sample space "$v$" we require $\sum_i P(s_i|v) = 1$, but this is not the case. The probabilities in Table 3 are not valid estimates of $P(s_i|v)$, nor do we have valid estimates of $P(s_i|\neg v)$.

---

[4] "question" = risk component

3. Calculation of likelihood ratios

The Risk Assessment User Manual says:

"*For each risk component, a likelihood ratio can be calculated, which can be thought of as the odds that a component will have a particular score, given that the species is invasive.*

$$L(s_i|v) = P(s_i|v)/P(s_i|\neg v) \tag{1}"$$

It is true that the RHS of equation (1) correctly specifies the likelihood ratios (as in Table 1), but valid estimates of neither $P(s_i|v)$ nor $P(s_i|\neg v)$ are available. Further, it is important to note that likelihood ratios *cannot* be thought of as conditional odds. This error in the analysis in the Risk Assessment Scheme arises because it has erroneously been assumed that $P(s_i|\neg v) + P(s_i|v) = 1$, so $P(s_i|\neg v) = 1 - P(s_i|v)$, and then $L(s_i|v) = P(s_i|v)/(1 - P(s_i|v)) = O(s_i|v)$. Of course, this is not the case.

4. Calculation of the combined odds

The Risk Assessment User Manual says:

"*The product of the likelihood ratios for all the components gives the combined odds that this set of scores will occur given that the species is invasive.*

$$O(s_{1..n}|v) = \prod_{i=1}^{i=n} L(s_i|v) \tag{2}"$$

It is true that when more than one risk component is available, the likelihood ratios for the available components can be multiplied together to provide a combined value (as mentioned in the discussion of the mammography experiment, see Go (1998)), provided the risk components are independent. However, as mentioned above, this value *cannot* be thought of as conditional odds. Looking at this from a different perspective, if we *did* have a set of odds and wanted to combine them, finding the product of the individual values would not provide a combined odds in a way that was consistent with the laws of probability. Also note, in passing, that the use of the index variable *i* for the product on the RHS of equation 2 is ambiguous, since *i* is used to index values of the score *s*. The product is to be taken over risk components, not scores.

5. Calculation of the odds that a species is invasive given the set of scores

The Risk Assessment User Manual says:

"*To obtain the odds that a species is invasive given the set of scores, it is necessary to consider the prior odds O(v) (i.e. the odds when no information is available) that a species is invasive.*

$$O(v|s_{1...n}) = O(v)O(s_{1..n}|v) \tag{3}"$$

The correct version of equation 3 is $O(v|\{s_i\}) = O(v) \times \prod_n L\{s_i\}$, in which the combined likelihood ratio is taken as the product of the individual likelihood ratios for *n* risk components: $\prod_n L\{s_i\} = L(s_i)_1 \times L(s_i)_2 \times \cdots \times L(s_i)_n$ and $\{s_i\}$ is the set of scores for the *n* risk components. Note that is not true in general to say that the prior odds is "the odds when no information is available". The prior odds represents the pre-test degree of belief that (in this case) a species is invasive. In fact, it is rather unlikely that no information at all is available before the Risk Assessment Scheme is invoked.

6. <u>The prior odds</u>

The Risk Assessment User Manual says:

"*In application of Bayes theorem to risk assessment in this way, the prior odds may not be particularly meaningful as, even if they could be calculated from a set of historical examples, they would depend on which species had been selected for investigation. The important thing is that the assessment is consistent and for a relative scoring system it is reasonable to arbitrarily set O(v)=1 (i.e. that, initially, there is a nominally equal chance that the species is invasive or not)*"

In choosing a prior odds $O(v)=1$, the authors are expressing a belief that species under consideration by the Risk Assessment Scheme have a prior probability of invasiveness $P(v)=0.5$. They are entitled to their belief, of course, but the justification offered is unconvincing. If the objective of the analysis is to formulate a "relative scoring system", then we note that the choice of value for $P(v)$ ($0<P(v)<1$) does not affect the relative ranking of species in terms of the posterior odds as calculated from equation 3 (either the incorrect or the correct version). Further, we might ask what the purpose of all this analysis is, if all that is required is a "relative scoring system". The original rating categories (Table 3) embodied a relative scoring system, and the score-to-probability conversion was carried out in such a way that the original score-based relative rankings would be preserved. The authors state that "*Score averaging often under-estimates high risk and over-estimates low risk. This project showed that the linear mapping of scores to conditional probabilities provides an important new approach to solve this problem.*" Unfortunately, if under- or over-estimation of risk is a real problem, the approach provided by the Risk Assessment Scheme not the solution.

7. <u>The posterior probability $P(v|s_i)$</u>

The Risk Assessment User Manual says:

"*Converting the odds (Eq. 3) back to a probability gives the final result, the conditional probability that a species is invasive given the set of scores obtained.*

$P(v|s_{1...n}) = O(v|s_{1...n})/(O(v|s_{1...n}) + 1)$ (4)"

As it stands, equation 4 is correct, but $P(v|s_{1...n})$ is meaningless as a "final result", considering the multiple errors involved in specifying $O(v|s_{1...n})$.

Finally we return, as promised, to the previously noted apparent contradiction that the probabilities in Table 3 are regarded both as $P(v|s_i)$ and as $P(s_i|v)$. We note now that the specific choice of $O(v)=1$ allows the authors to consider their equation 3 as:

$O(v|s_{1...n}) = O(s_{1..n}|v)$

in which case P($v|s_{1...n}$) = P($s_{1...n}|v$), so providing a justification for the reversal of conditionality without further ado. This reasoning is, of course, entirely dependent on the specific choice of prior probability P($v$)=0.5 and on the erroneous form of equation 3.

**Conclusion**

The UK Non-Native Organism Risk Assessment Scheme does not provide a valid analysis on which to base predictive discrimination between invasive species, likely to cause environmental problems if imported, and non-invasive species, unlikely so to do. For further details of the errors in the analysis of conditional probability, based on the application of Bayes' theorem, see Hughes (2008).